\documentclass[5p]{elsarticle}

\usepackage{lineno}
\usepackage[numbers]{natbib}
\usepackage{placeins}
\usepackage{xcolor}
\usepackage{float}
\usepackage{mathtools,hyperref}
\usepackage[T1]{fontenc}
\usepackage{multirow}

\restylefloat{table}

\restylefloat{figure}

\setcitestyle{authoryear,open={(},close={)}} 
\journal{XXX}

\bibstyle{elsarticle-num}

\begin{document}

\begin{frontmatter}

\title{Predicting Mutual Funds' Performance using Deep Learning and Ensemble Techniques}

\author{Nghia Chu}\fnref{myfootnote1}
\address{Vega Corporation, Tang 3, 106 Hoang Quoc Viet, Nghia Do, Cau Giay, Hanoi, Vietnam \\ cdnghia79@gmail.com}
\fntext[myfootnote1]{Corresponding Author}

\author{Binh Dao}
\address{Hanoi University, Km 9, Nguyen Trai, Nam Tu Liem, Hanoi, Vietnam \\ binhdtt@hanu.edu.vn}

\author{Nga Pham}
\address{Monash Centre for Financial Studies, 30 Collins St, Melbourne, VIC 3000, Australia \\ nga.pham@monash.edu }

\author{Huy Nguyen}
\address{The University of Melbourne, Melbourne Connect
700 Swanston St, Carlton, VIC 3053, Australia
 \\ huyxuan.nguyen@student.unimelb.edu.au}

\author{Hien Tran}
\address{Hanoi University, Km 9, Nguyen Trai, Nam Tu Liem, Hanoi, Vietnam \\ hientttfmt@hanu.edu.vn}

\begin{abstract}
Predicting fund performance is beneficial to both investors and fund managers and yet is a challenging task. In this paper, we have tested whether deep learning models can predict fund performance more accurately than traditional statistical techniques. Fund performance is typically evaluated by the Sharpe ratio, which represents the risk-adjusted performance to ensure meaningful comparability across funds. We calculated the annualized Sharpe ratios based on the monthly returns time series data for more than 600 open-end mutual funds investing in listed large-cap equities in the United States. With five different forecast horizons ranging from 6 to 24 months, we find that long short-term memory (LSTM) and gated recurrent units (GRU) deep learning methods, both trained using modern Bayesian optimization, provide higher accuracy in forecasting funds' Sharpe ratios than traditional statistical ones. An ensemble method which combines forecasts from LSTM and GRU achieves the best performance of all models. There is evidence to say that deep learning and ensembling offer promising solutions in addressing the challenge of fund performance forecasting. 

\end{abstract}


\end{frontmatter}


\section{Introduction}
\label{sec:sample1}
Investment funds are important intermediaries of the world's financial market. Globally, there were more than 126,000 regulated mutual funds in 2020. In the United States alone, there were around 9,000 mutual funds with 24 trillion dollars of assets under management (AUM) in 2020, according to the Investment Company Fact Book by \citep{InvestmentCompanyFACTBOOK}. Choosing the right investment fund has always been challenging for institutional and individual investors. Fund performance is driven by various factors, including the fund managers' understanding of the macro environment, industry factors, stock performance, and trading skills. While it is always written in most funds' prospectus that past performance is not indicative of future performance, there has been ample empirical literature that documents fund performance persistence, for example, among fixed income funds as in \citep{kahn1995does}, equity funds and real estate funds \citep{arnold2019private}. Some attribute performance persistence of mutual funds to managers' skills \citep{BERK20151, KACPERCZYK2014JF} whereas others explain performance persistence by common factors affecting stock returns and persistent differences in fund expenses and transaction costs \citep{Carhart1997} and capital flows \citep{Lou2012}.

While the asset pricing literature centers around factors that explain fund performance and performance predictability, the topic of forecasting fund performance seems to be under-explored. It is, therefore, interesting to answer whether future fund performance can be forecast given past performance data, especially with the advanced techniques in deep learning for time-series forecasting.

Being able to forecast fund performance with confidence will allow asset owners, typically individual investors, endowments, foundations, pension funds and sovereign funds, to make more informed decisions about their future asset manager selection. It is well documented in the literature that investors recognize better-skilled managers and reward outperformance with more fund flows \citep{BERK20151}. Better performance predictability will drive fund flows in the market to outperforming funds, at the same time, creates pressure for underperforming funds to improve. This process enhances market efficiency. 

Despite its significance, there has been a lack of research in forecasting fund performances, especially research that involves forecasting the performance of multiple mutual funds as multiple simultaneous time series using one model. This research is intended to fill this gap. We aim to evaluate various techniques, including modern deep learning algorithms, traditional statistical techniques, and ensemble methods that combine model forecasts in different ways. 

This work investigates whether deep learning techniques can better forecast future fund performances based on past performance data than conventional statistical methods. Deep learning comprises sophisticated neural network-based methods designed to learn patterns from data. Deep learning has demonstrated its effectiveness in many prediction problems that involve natural language and vision datasets. While there have been certain advances for time series data, there is still a lack of convincing evidence that deep learning outperforms other simpler, statistics-based methods on different problem domains. 

Past research concluded that neural networks were not well suited for time series forecasting \citep{hyndman2018brief}. This may be because numerous deep learning-based algorithms require large amounts of data to perform well. For example, the famous BERT model, a transformer used for natural language processing, was pre-trained on an extremely large text corpus, including the English Wikipedia with over 2,500M words \citep{kenton2019bert}. The nature of time series data does not permit such a massive amount of data available for training deep learning models. 

Contrarily, recent research in deep learning applied to the time series domain shows more promising results. In the M4 forecasting competition, the hybrid approach integrating exponential smoothing into advanced neural networks outperforms other statistical methods and wins the competition as in \citep{smyl2020hybrid}. \citep{hewamalage2021recurrent} investigates the effectiveness of recurrent neural networks in forecasting various datasets. The study concludes that RNNs are competitive alternatives to state-of-the-art univariate benchmark methods such as ARIMA and exponential smoothing (ETS). 

These studies have in common that they investigate the recurrent neural network (RNN) algorithms that are well-known approaches in sequence modelling. They also develop global deep learning models capable of learning from multiple time series to improve prediction accuracy for individual ones. This technique is known as cross-learning, which has been found as a promising alternative to traditional forecasting \citep{semenoglou2021investigating}. 

Our study addresses the question of whether cross-learning with carefully tuned RNNs, specifically LSTM and GRU, works more effectively than traditional univariate methods in the problem of forecasting performances for multiple mutual funds. The sample of mutual funds is relatively homogeneous in that they adopt a similar investment strategy in the United States, which increases the chance that information in one time series would help predict the future values of other time series. 

We compare the accuracy of ARIMA, Theta, ETS and Naive statistical methods with deep learning algorithms in predicting fund performance. While ARIMA and ETS are popular univariate statistical forecasting models, Theta was the best-performing forecasting algorithm in the M3 competition, according to \citep{makridakis2000m3}. The Naive method is included for comparison as, despite its simplicity, it works well for various financial and economic time series \citep{hyndman2018forecasting}. 
In addition to individual deep learning and statistical algorithms, we investigate how various ensembling methods for combining model forecasts lead to different performance gains. 

In summary, the contributions of our research are:
\begin{itemize}
  \item To the best of our knowledge, our research is the first that addresses the problem of forecasting the performance of multiple mutual funds simultaneously using modern deep learning approaches. 
  \item We compare deep learning performances against well-known statistical methods using rolling origin cross-validation with well-defined data splitting methods and multiple forecast horizons.
  \item We experiment with different ensembling methods for combining model forecasts, including local out-of-sample performance per time series per algorithm and global out-of-sample performance per algorithm in computing weighted averages of forecasts. 
  \item We present a detailed methodology for training deep learning models, specifically for LSTM and GRU algorithms, using modern Bayesian optimization. We believe the methodology benefits machine learning practitioners aiming to study the hyper-parameter optimization process of deep learning models.
\end{itemize}

\section{Study background}
This section presents the literature on predicting the performance of mutual funds, traditional univariate forecasting models, and deep learning models for time series forecasting. 

\subsection{Predicting mutual fund performance}
Fund performance is fundamentally assessed based on two aspects (1) expected return and (2) risk level, i.e., the variability of returns, measured by the standard deviation of return. Although the expected rate of return of a fund is a critical indicator, it alone does not provide an accurate picture of how the fund has performed relating to other funds, as higher expected returns come at the cost of having a higher level of risk. Different funds could exhibit different variability of returns due to differences in risk-taking strategies, diversification levels or management skills. 

Since the 1960s, the Sharpe ratio introduced by \citep{sharpe1966} and modified in \citep{SharpeWilliamF1994TSR} has become the most popular performance statistics of mutual funds for considering the trade-off between fund's risk and return. Sharpe ratio is also known as the reward-to-variability ratio, as a fund's excess return (the rate of return over and above the risk-free rate of return) is assessed against its level of risk, i.e., the variability of returns. Therefore, the Sharpe ratio is unaffected by scale. The implied assumption of the Sharpe ratio is that all investors can invest funds at the risk-free interest rate, representing the opportunity cost of investing in an investment fund or the cost of capital invested. Fund managers are under constant pressure to deliver a high Sharpe ratio, otherwise risking losing capital allocated to their fund.

While other performance indicators are introduced in academic literature, the Sharpe ratio has always been among the preferred choices of industry practitioners when it comes to fund performance evaluation, thanks to its simplicity. It is reported by most asset managers and included in large fund databases such as Morningstar or Lipper, cited reports from \citep{EltonEdwinJ2020ARot}. Therefore, in this paper, we use the Sharpe ratio as an indicator of fund performance.

Given the scant literature on forecasting fund performance, the selection of a reasonable forecast horizon remains an open question. Our considerations include a wide range of time windows in which performance persistence is found to hold, as identified in the persistence literature stream. Persistence studies investigate whether funds' relative performance is sustained over adjacent periods. Some find persistence to be short-lived \citep{BollenNicolasP.B.2005SPiM}, with performance persistence observed only within six months \citep{KACPERCZYK2014JF} or one year \citep{Carhart1997}. \citep{Lou2012} find that evidence for funds' persistence reversed after six to twelve quarters. However, \citep{Irvine2022} provide evidence that outperformance persists for at least 17 to 24 months, depending on different asset pricing models. For longer time windows, \citeauthor{BERK20151} report large cross-sectional performance differences in managers' skills persist for as long as ten years. Based on the diverse predictability and persistence time windows observed in the literature, we decided to use forecast horizons of 6, 9, 12, 18 and 24 months, with funds' performance being forecasted for every single month during a forecast horizon.

Most studies on forecasting fund performance have used traditional time series forecasting models. Starting from the 1980s and 1990s, more advanced neural network techniques were used to address financial forecasting questions. Specifically, the use of neural networks to forecast fund performance was introduced in the papers of \citep{CHIANG1996205} and \citep{IndroANN1999}. \citep{CHIANG1996205} employed neural networks to forecast mutual fund net asset value, while \citep{IndroANN1999} used the same approach to forecasting mutual fund performance in terms of percentage returns. 

A more recent paper is  \citep{WangKehluh2010Ufan}. While \citep{WangKehluh2010Ufan} also use the Sharpe ratio to measure fund performance like ours, they compare the fast adaptive neural network classifier (FANNC) with the Back Propagation neural network (BPN) model and report that the FANNC is more efficient than the BPN in both classifying and predicting the performance of mutual funds.\\
It should also be noted that \citep{WangKehluh2010Ufan} use manager’s momentum strategies and herding behavior as input variables for the prediction of funds' Sharpe ratio whereas, in our approach, past Sharpe ratios are the input to forecast future ones.

The study of \citep{pan2019prediction} is similar to the paper of \citep{WangKehluh2010Ufan} in which they use three inputs of fund size, annual management fee and fund custodian fee and the fund return as the output for data envelopment analysis, Back Propagation Neural Network and GABPN (new evolutionary method) for 17 open-end balance stock funds for the period August 31, 2015 to July 1, 2016. This study also uses five accuracy measures to evaluate forecast performance, analyzes the rate of return and builds the mutual fund net worth prediction model. Compared to \citep{pan2019prediction}, our paper examined a substantially larger number of funds over a much longer time window using extensive forecasting methods such as deep learning, ensemble, and traditional statistical techniques.

\subsection{Traditional univariate forecasting models}

Time series forecasting, especially stock price and fund performance forecasting, have traditionally been popular research topics in statistics and econometrics. Traditional methods range from simple seasonal decomposition (STL) and simple exponential smoothing (SEM) to more complex ones such as ETS \citep{hyndman2008forecasting} and ARIMA (\citeauthor{newbold1983arima}, \citeyear{newbold1983arima}; \citeauthor{box2015time},\citeyear{box2015time}). Non-linear autoregressive time series using numerical forecasting procedure (m-step-ahead predictive) is also used in \citep{cai2005forecasting} and automatic ARIMA in  \citep{melard2000automatic} using Time Series Expert TSE-AX.
Combining several traditional methods can be seen in \citep{bergmeir2012use}, which used six traditional forecasting methods, namely autoregression (AR), moving average (MA), combinations of those models (ARMA), and integrated ARMA as ARIMA, Seasonal and Trend decomposition using Loess (STL) as well as TAR (Threshold AR) based on cross-validation and evaluation using the series’ last part. They suggest that the use of a blocked form of cross-validation for time series evaluation should be the standard procedure.

The traditional uni-variate methods dominated other complex computational methods at many forecasting competitions, including M2, M3, and M4 as in \citep{ahmed2010empirical}; \citep{crone2011advances}; \citep{makridakis2000m3}. A recent paper, \citep{fiorucci2020groec}, proposes a methodology for combining time series forecast models. The authors use a cross-validation scheme that assigns higher weights to methods with more accurate in-sample predictions. The methodology is used to combine forecasts named generalized rolling origin evaluation from traditional forecasting methods such as the Theta, Simple Exponential Smoothing (SES), and ARIMA models. \citep{makridakis2020m4} reported 61 forecasting methods used in M4 competition with 100,000 time series.
The paper of \citep{li2020forecasting} used the ARIMA model with deep learning to forecast high-frequency financial time series. 

The traditional uni-variate methods show that they work well when the data volume is not extensive (\citealp{bandara2020forecasting}; \citealp{xu2019hybrid} and \citealp{sezer2020financial}) and a low number of parameters to be estimated.

\subsection{Deep learning forecasting models}
Recently, several papers combine the traditional methods with deep learning  (\citealp{li2020forecasting}; \citep{pannakkong2017novel} and \citealp{xu2019hybrid}). \citep{li2020forecasting} work on the Chinese CSI 300 index and compare method as Monte Carlo numerical simulation, ARIMA, support vector machine (SVM), long short-term memory (LSTM) and ARIMA-SVM models. Their results show that the enhanced ARIMA model based on LSTM not only improves the forecasting accuracy of the single ARIMA model in both fitting and forecasting but also reduces the computational complexity of only a single deep learning model. \citep{pannakkong2017novel} proposes a hybrid forecasting model involving autoregressive integrated moving average
(ARIMA), artificial neural networks (ANNs) and k-means clustering. \citep{xu2019hybrid} use all linear models and deep belief network (DBN) models. 

Several other papers, \citep{jianwei2019novel},  \citep{cao2019stock}, and \citep{sezer2020financial} use mixture models for financial commodity and stock forecasting. \citep{jianwei2019novel} employs independent component analysis (ICA), gate recurrent unit neural network (GRUNN) named ICA-GRUNN among others to estimate gold prices. The paper shows that ICA-GRUNN produces prediction with high accuracy and outperforms the benchmark methods, namely ARIMA, radial basis function neural network (RBFNN), long short-term memory neural network (LSTM), GRUNN and ICA-LSTM.
\citep{cao2019stock} proposes the convolutional neural network (CNN) and CNN-support vector machine (SVM) to forecast stock index. The paper concludes that the neural network method for financial prediction can handle continuous and categorical variables and yield good prediction accuracy.
\citep{sezer2020financial} presents a review of financial time series forecasting methods. Machine Learning (ML) models (Artificial Neural Networks (ANNs), Evolutionary Computations (ECs), Genetic Programming (GP), and Agent-based models), Deep Learning (DL) models (Convolutional Neural Networks (CNNs), Deep Belief Networks (DBNs), and Long-Short Term Memory (LSTM)) have appeared within the field, with results that significantly outperform their traditional ML counterparts for data like market indices, foreign exchange rates, and commodity prices forecasting.
\citep{hewamalage2021recurrent} provides a big picture of the perspective direction for time series forecasting using recurrent neural networks. The authors present an extensive study and provide a software framework of existing RNN architectures for forecasting as Stacked Architecture and Sequences to Sequence (S2S) for three RNN units including Elman RNN, LSTM and GRU. The main difference between S2S and Stacked with the dense layer and the moving window input format is that in the former, the error is calculated per each time step and the latter calculates the error only for the last time step.

In \citep{chen2021machine}, a multivariate model is proposed to explain Bitcoin price with other economic and technology variables, using both traditional statistical and deep learning models. They use RNN (Recurrent Neural Network) and RF (Random Forest) for the first stage to detect the importance of input variables. They propose a wide range of models for the second stage such as ARIMA (Autoregressive Integrated Moving Average), SVR (Support Vector Regression), GA (Genetic Algorithm), LSTM (Long Short-Term Memory), ANFIS (Adaptive Network Fuzzy Inference System), for bitcoin forecast in different periods.\\

Our research fills the gap of the lack of modern deep learning research for predicting funds' risk-adjusted performance measured by the Sharpe ratio. The results of our research show that deep learning and ensembling are promising approaches to addressing the challenge of predicting mutual funds' performance.
 
\section{Methodology}

In this section, we present the methodology employed for this research, from cross-validation scheme, model selection to model training and other techniques to ensure the robustness of our results.  

\subsection{Cross-validation scheme}
\label{cv-scheme}
Cross-validation has been widely used in machine learning to estimate models' prediction errors as described in \citep{ruppert2004elements}. This method estimates the mean out-of-sample error, also known as the average generalization error by \citep{ruppert2004elements}.

For normal machine learning problems, where there is no time dimension, k-fold cross-validation is the common cross-validation technique applied in practice. This technique requires the dataset to be randomly divided into a train set and a validation set, in k different settings. The random division makes each element of the dataset exist in the validation set exactly once and in the training set at least once. 

In time series forecasting, there is a time dimension, and care must be taken to ensure that time-dependent characteristics are preserved during dataset setup for model evaluations. For example, we cannot simply and randomly assign an element to the validation set, since this may interrupt time order between observations and lead to data leakage, where future data may be used to predict past data.

The study of \citep{bergmeir2012use} showed that the use of cross-validation in time series forecasting led to a more robust model selection. In this research, we follow the rolling origin scheme to divide our dataset into six cross-validation splits. Our choice of six cross-validation splits is identical to the study of \citep{fiorucci2020groec}. To limit the computational costs when training deep learning models within reasonable limits, it is beyond the scope of our study to evaluate the effects of different cross-validation splits on the study results. 

Models are trained on each training set and then evaluated on the respective validation set. The error measured in each validation set is the estimated generalization error for that particular fold. Averaging the generalization error estimated for all the folds will yield the estimated generalization error of the model. Model selection is then carried out by comparing average generalization errors across all the models trained. 

Specifically, our cross-validation scheme is a slightly adjusted version of Generalized Rolling Origin Evaluation (GROE), as described in \citep{fiorucci2020groec}. Below are the properties of this GROE scheme in connection to our study:
\begin{itemize}
    \item $p$ denotes the number of origins, where an origin is the index of the last observation of a train set. The origins are referred to as $n_1, n_2, . . . , n_p$, which can be found recursively through the equation
$n_{i+1} =n_i+m$. In this equation, $m$ represents the number of data points between two consecutive origins. In our study, the number of origins $p$ equals six, which means original data are split in six different ways to produce six train/validation set pairs.
    \item $H$ represents the different forecast horizons, which are 6, 9, 12, 18, and 24 months in our study.
    \item $n$ is the length of each time series and $n_1$ is equal to $n - H$ if $n - H$ is greater than or equal a certain threshold.
    \item $m$ can be calculated as the biggest integer lower than or equal to $H/p$, which is problem dependent. 
\end{itemize}

We recognize that if we strictly follow GROE, this may lead to validation sets with different sizes. We believe that setting the length of all the validation sets equal and equal to the forecast horizon will make the validation sets represent the actual forecast dataset when applying a forecasting model in practice better. Therefore, we adjust the method slightly by first determining the last train/validation sets, and then working backwards to determine the other train/validation sets. In our method, the last validation set is defined to consist of the last $H$ observations in the dataset, and the corresponding train set consists of all the observations in the dataset prior to the validation set. Then we shift the origin of the last train set, which is $n - H$, back in time $m$ data points to find the origin of the previous train sets, i.e. using $n_i = n_{i+1} - m$. 

This modification ensures all validation sets have the same length, which is exactly equal to the forecast horizon. As an example, for the forecast horizon of 18 months, our cross-validation scheme results in the following division of train and validation sets:

\begin{figure}[H]
    \centering
    \includegraphics[scale=0.8]{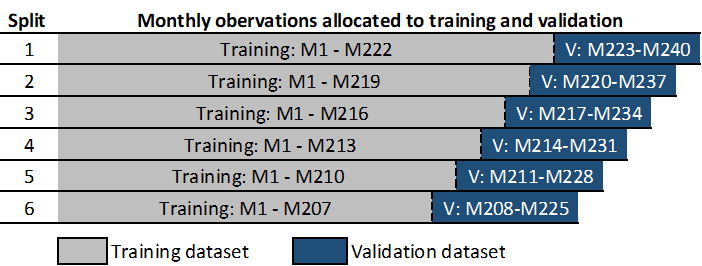}
    \caption{Train and validation sets for a forecast horizon of 18m} 
    \label{fig:TrainValidation}
\end{figure} 

\subsection{Recurrent neural networks}
In this research, we employ two modern types of recurrent neural networks (RNNs) known as LSTM and GRU and compare their effectiveness in forecasting fund performances against their statistical counterparts. In this section, we present the unit design and mathematics fundamentals of these techniques. In addition to LSTM and GRU, we also describe the vanilla RNNs which provide the foundation for more advanced RNN architectures.

\subsubsection{Vanilla recurrent neural networks}

\begin{figure}[H]
    \centering
    \includegraphics[scale=0.8]{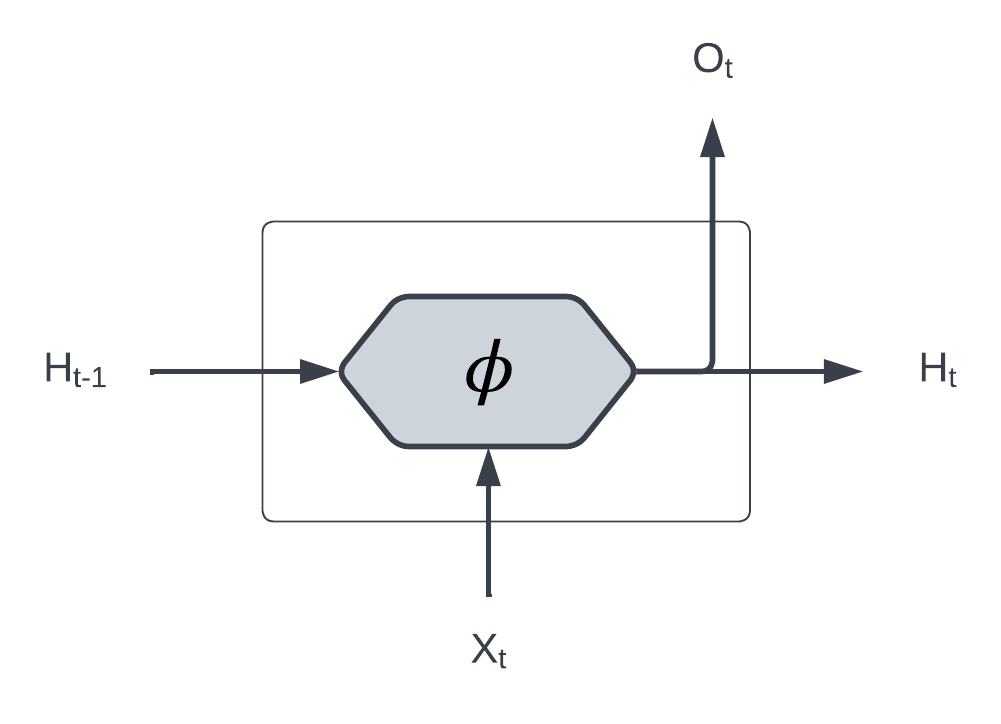}
    \caption{A Basic Recurrent Cell}  
    \label{fig:RNN} 
\end{figure}

RNNs are designed to address the problem of learning from and predicting sequences. They have been widely applied in the field of natural language processing. In the following figure and equations, we illustrate the internal workings of RNNs. The basic RNN cell has the structure as shown in Figure \ref{fig:RNN}.

\begin{equation} \label{equa1}
H_{t} = \phi \left( H_{t-1} \cdot W_{h1} + X_{t} \cdot W_{h2} + b_{h} \right)
\end{equation}

\begin{equation}
\label{equa_2}
O_{t}=H_{t} \cdot W_{o}  + b_{o} 
\end{equation}

In equation \ref{equa1}, $H_{t}$ denotes the hidden state of the RNN cell, and $X_{t}$ denotes the input of the cell at the current time step (i.e. time step $t$). $W_{h1}$ and $W_{h2}$ are matrices of weights whereas $b_{h}$ denotes the bias vector for the hidden state. $\phi$ is the activation function of the hidden state. This equation describes the recurrent computation which gives rise to the term "recurrent", where the current hidden state $H_{t}$ is computed based on the previous hidden state  $H_{t-1}$ and the current input $X_{t}$. This recurrence in computation leads to the ability of the hidden state $H_{t}$ to store information of the sequence up to time step $t$. In a similar way, the hidden state of the next time step $H_{t+1}$ is computed based on  $H_{t}$ and the input $X_{t+1}$ at the next time step. 

Equation \ref{equa_2} shows how the output is computed using the hidden state of the current time step. In equation \ref{equa_2},  $O_{t}$ denotes the output of the cell at time step $t$, $W_{o}$ is the weight matrix, and $b_{o}$ represents the bias vector for the cell output.  

Even though the recurrent computation makes it possible for RNNs to carry past information into the current time step, RNNs has limited capability in handling long-term dependencies in sequential data. This type of network suffers from the problems of vanishing and exploding gradients, making learning in long sequences difficult. Vanishing gradients happen when gradients during backpropagation become vanishingly small, and the weights cannot be updated adequately, whereas exploding gradients occur when large gradients accumulate during backpropagation which results in unstable models as model weights receive very large updates.

\subsubsection{Long short-term memory}
Long short-term memory (LSTM), introduced in \citep{hochreiter1997long} extends the capacity of vanilla RNNs to be able to remember and effectively handle longer sequences. The computation of the hidden state in an LSTM is illustrated in Figure \ref{fig:LSTM}:
\begin{figure}[H]
    \centering
    \includegraphics[scale=0.6]{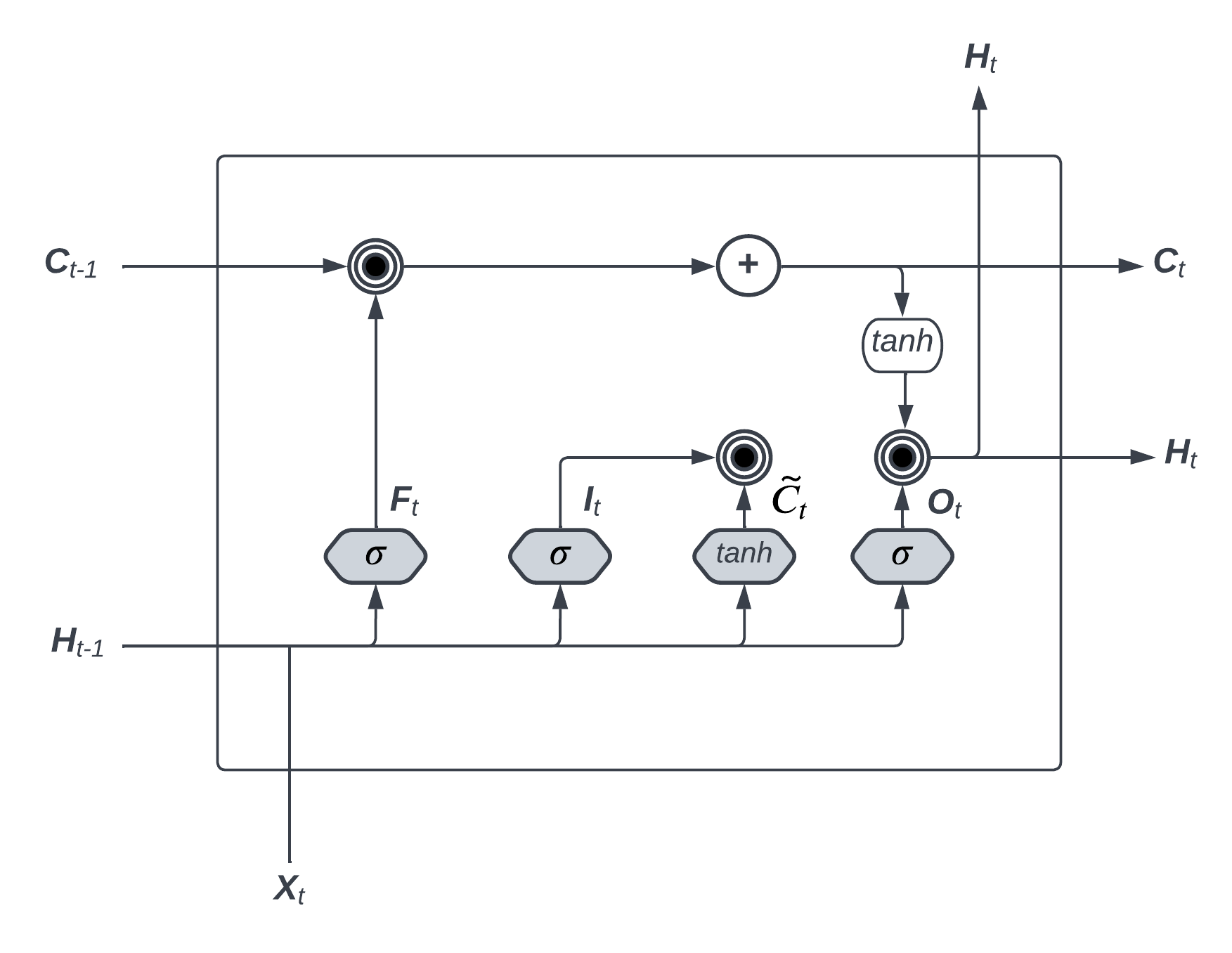}
    
   \caption{Computation of hidden state in LSTM} 
    \label{fig:LSTM}
\end{figure}

In Figure \ref{fig:LSTM}, there are three gates designed to regulate information flow through the memory cell. Specifically, the input gate controls how much information to read into the cell, the forget gate decides how much of the past information to forget and how much to retain, and the output gate reads output from the cell. This design enables the cell to decide when to remember and when to ignore inputs, which is essential in remembering useful information and forgetting less useful one. 

Mathematically, these gates are computed as follows:

\begin{center}
$I_{t}=\sigma\left(H_{t-1} \cdot W_{i1}  + X_{t} \cdot W_{i2} + b_{i}\right)$
$F_{t}=\sigma\left( H_{t-1} \cdot W_{f1} +  X_{t} \cdot W_{f2} + b_{f}\right)$
$O_{t}=\sigma\left(H_{t-1} \cdot W_{o1} + X_{t} \cdot W_{o2} + b_{o}\right)$
\end{center}

In the above equations, $I_{t}$, $F_{t}$ and $O_{t}$ represent the input gate, forget gate and output gate at the current time step $t$. $X_{t}$ is the input at time $t$ and $H_{t-1}$ is the hidden state at the previous time step. The $W_{s}$ represent the weight matrices and the $b_{s}$ are bias vectors. $\sigma$ is the sigmoid activation function that makes the values of these gates in the range of (0, 1).

The memory cell and the hidden state are computed as follows:
\begin{center}
$\tilde{C}_{t}=\tanh \left(H_{t-1} \cdot W_{c1}  + X_{t} \cdot W_{c2} + b_{c}\right)$\\
$C_{t}=F_{t} \odot C_{t-1} + I_{t} \odot \left(\tilde{C}_{t} \right)$\\
$H_{t}=O_{t} \odot \tanh \left(C_{t}\right)$\\
\end{center}

In the above equations, $\tilde{C}_{t}$ represents the candidate memory cell and $C_{t}$ represent memory cell content. Similar to the equations of the gates, $W_{s}$ represent the weight matrices and the $b_{s}$ are bias vectors. The computation of the candidate memory cell is similar to those of the gates, except that it uses the tanh activation function instead of sigmoid. 

The candidate memory cell captures both past information from $H_{t-1}$ and information from the current input $X_{t}$. The computation of memory cell content $C_{t}$ is based on past memory cell state represented by  $C_{t-1}$ and the current memory cell candidate which serves as current input. The element-wise matrix multiplication denoted by $\odot$ makes it possible for controlling how much of past information is forgotten (by multiplying $F_{t}$ by $C_{t-1}$) and how much of current input is retained (by multiplying $I_{t}$ by $\tilde{C}_{t}$). Both $F_{t}$ and $I_{t}$ use sigmoid as the activation function, which produces values in the range (0, 1) to control how much information is discarded/retained in element-wise matrix multiplication. For example, when values of $F_{t}$ are close to zero, the result of element-wise matrix multiplication between $F_{t}$ by $C_{t-1}$ will make past information close to zero or in other words past information become forgotten. 

The computation of the hidden state $H_{t}$ depends on the memory cell $C_{t}$ and how much of the memory cell is passed as output is controlled by the output gate $O_{t}$. 

Our study uses the above described design for LSTM, in contrast to \citep{hewamalage2021recurrent} which uses LSTM with peephole connections. We also use the stacked architectures for both LSTM and GRU, where the term ``stacked" means that different recurrent layers can be stacked on top of each other.   

\subsubsection{Gated recurrent units}
Gated recurrent units (GRU) \citep{cho2014properties} were introduced almost two decades after LSTM. GRU possess a similar but simpler architecture than that of LSTM, which gives them the advantage of faster computation. GRU also use gates to regulate information flow, but it contains only the reset gate and the update gate.

\begin{figure}[H]
    \centering
    \includegraphics[scale=0.6]{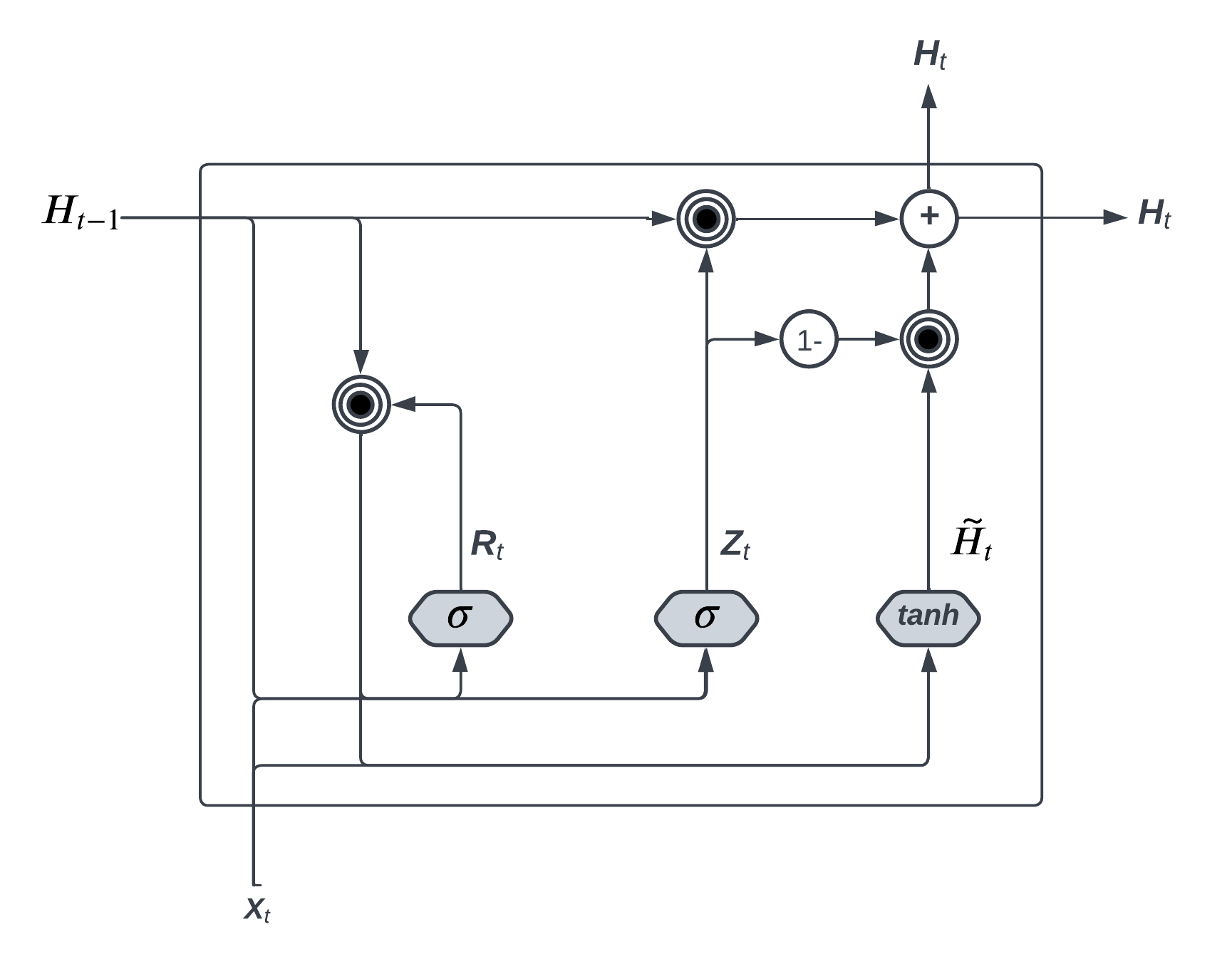}
   \caption{Computation of hidden state in GRU} 
    \label{fig:GRU}
\end{figure}

The computation of the hidden state in GRU is described in Figure \ref{fig:GRU}.

\begin{center}
\begin{align*}
Z_{t}&=\sigma\left(H_{t-1} \cdot W_{z1} + X_{t} \cdot  W_{z2} + b_{z}\right) \\
R_{t}&=\sigma\left(H_{t-1} \cdot W_{r1} + X_{t} \cdot W_{r2} + b_{r}\right)\\
\end{align*}  
\end{center}

The above two equations show the mathematical mechanisms of the two gates in GRU: update gate ($Z_{t}$) and reset gate ($R_{t}$). The computation of both the update gate and reset gate at the current time step $t$ is based on the hidden state at the previous time step $H_{t-1}$ and the current input $X_{t}$. The $W_{s}$ are the weight matrices and $b_{s}$ are the bias vectors. $\sigma$ is the sigmoid activation function that makes the values of these gates in the range of (0, 1). 

The next equations show the computation of the candidate hidden state ($\tilde{H}_{t}$) and the hidden state ($H_{t}$).
\begin{center}
\begin{align*}
\tilde{H}_{t}&=\tanh \left( (R_{t} \odot H_{t-1}) \cdot W_{h1} + X_{t} \cdot W_{h2}  + b_{h}\right) \\
H_{t}&=(1-Z_{t}) \odot \tilde{H}_{t} + Z_{t} \odot H_{t-1}      
\end{align*}
\end{center}

The candidate hidden state at the current time step $\tilde{H}_{t}$ captures both past information from the previous hidden state $H_{t-1}$ and information from the current input $X_{t}$. The element-wise matrix multiplication of $H_{t-1}$ and $R_{t}$ makes it possible for the reset gate to control how much of past information in $H_{t-1}$ is retained as the value of $R_{t}$ is in the range (0, 1). If, for example, $R_{t}$ equals one, all of the information of $H_{t-1}$ is retained, and if $R_{t}$ equals zero, all previously hidden state information is discarded. The candidate hidden state, therefore, contains the current input information plus parts of the previously hidden state information as controlled by the reset gate. 

In the next equation, the current hidden state $H_{t}$ is computed from the current candidate hidden state $\tilde{H}_{t}$ and the hidden state from the previous time step $H_{t-1}$. The update gate is used to control how much information from these two components is passed to the current hidden state. When the update gate $Z_{t}$ is close to one, the current hidden state takes most of the information from the previous hidden state whereas when the update gate $Z_{t}$ is close to zero, the current hidden state takes most of the information from the current candidate hidden state.

A previous study conducted to evaluate GRU and LSTM algorithms on the sequence modelling tasks of polyphonic music modelling and speech signal modelling found GRU to be comparable to LSTM \citep{chung2014empirical}. However, it is unclear which type of these algorithms performs better in forecasting fund performance. Therefore, our research aims to evaluate and compare these two deep learning methods in the context of fund performance forecasting. Our study implements these methods using the Keras and Tensorflow libraries in Python where Bayesian optimization is used for hyper-parameter optimization, which will be described in a later section. 

\subsubsection{Ensemble methods}
Ensemble methods are ways to combine forecasts from different models with the aim to improve models' performance. There are different methods to combine forecasts. The simplest method would be averaging the model predictions for different algorithms. Other methods can make use of weighted averages. The study of \citep{fiorucci2020groec} combines time series forecast models with the weights proportional to the quality of each model's in-sample predictions in a cross-validation process. Their findings reveal that forecast results improve compared to methods that use equal weights for combination. However, all of the methods used for combination in their study are statistical methods, and in-sample prediction qualities are used to determine weights. 

Our approach in ensembling forecast results is different from the above study in that in addition to combining forecasts of top-performing models, we also combine forecasts for models belonging to different approaches, i.e. all deep learning models combined with all statistical models. According to \citep{petropoulos2022forecasting}, combined forecasts work more effectively if the methods that generate forecasts are diverse, which leads to fewer correlated errors. Another difference is that we use out-of-sample model prediction qualities to determine weights, which we believe are more objectively representative of models' performance than using in-sample prediction qualities. The reason is that deep learning models may easily overfit the training data and produce high-quality in-sample predictions while the corresponding performances can significantly drop on out-of-sample data. 

In this study, we compare three ensembling methods:
\begin{itemize}
  \item The first method uses equal weights in averaging model forecasts. This is the simple form for combining forecasts, known as simple averages.
  \item The second method calculates a weighted average of model forecasts where the weights are \textit{inversely proportional} to the final out-of-sample MASE (our main forecasting performance metric used for evaluation of forecasting models) result of each algorithm. The weights represent the overall (or global) average out-of-sample model quality across all time series for each algorithm. 
  \item The third method calculates a weighted average of model forecasts where the weights are \textit{inversely proportional} to the average out-of-sample MASE result of each algorithm for the particular time series in question. The weights represent the specific (or local) average out-of-sample model quality for each time series in each algorithm. 
 \end{itemize}

\subsection{Traditional statistical models for time series forecasting}
 To compare against deep learning algorithms, we select three well-known traditional statistical methods for time series forecasting, which are ARIMA, ETS, and Theta, as well as the Naive method, which is known to work well for numerous financial and economic time series. The Theta method of forecasting, introduced by \citep{assimakopoulos2000theta}, is a special case of simple exponential smoothing with drift. These four methods are recommended by \citep{petropoulos2022forecasting} when benchmarking new forecasting methods. Further detail about the four models can be seen in \citep{hyndman2018forecasting}. 
 
 In our experiments, we use an optimized version of the Theta model proposed by \citep{fiorucci2016models}. We use the well-known \textit{forecast} R package introduced in \citep{hyndman2008automatic} for training ETS, ARIMA and Naive models, and the \textit{forecTheta} R package for training the optimized Theta models. These models are trained using the parallel processing capabilities provided in the \textit{furrr} R package. 
 
\section{Experiment setup}

\label{sec:sample2}
\subsection{Data description}
Our dataset started with the monthly returns of more than 1200 mutual funds investing in listed large-cap equities in the United States available on Morningstar Direct. Daily returns series are not available. We required at least 20 years of data up to October 2021 and obtained 634 funds. These funds are relatively homogeneous as they share the same investment strategy and are all domiciled in the United States. From the monthly returns series, we computed the annualised Sharpe ratios for each fund on a rolling basis. Table 1 summarises the statistics of these time series.\\

\begin{table}[ht]
\small
    \centering
    \scalebox{0.85} {
    \begin{tabular}{|l|l|r|r|r|r|}
\hline  && \text { Mean } & 
\text {Sd } 
& \text { Min } & \text { Max } \\
\hline {\text { us45890c7395}}  & \text { Return } & 4.351	&	27.069	&	-129.960	&	62.540 \\
\cline { 2 - 6 } & \text { Sharpe } & 0.213	&	0.542	&	-1.444	&	2.132 \\

\hline {\text { us5529815731}} & \text { Return } & 7.522	&	23.741	&	-97.820	&	70.040 \\
\cline { 2 - 6 } & \text { Sharpe } & 0.295	&	0.555	&	-1.430	&	1.790 \\

\hline {\text { us82301q7759}} & \text { Return } & 8.140	&	22.494	&	-100.080	&	69.620 \\
\cline { 2 - 6 } & \text { Sharpe } & 0.301	&	0.571	&	-1.444	&	2.507\\

\hline {\text { us6409175063}} & \text { Return } & 8.804	&	29.045	&	-114.100	&	93.900 \\
\cline { 2 - 6 } & \text { Sharpe } & 0.304	&	0.584	&	-1.288	&	3.129 \\

\hline {\text { us1253254072}} & \text { Return } & 15.065	&	35.222	&	-156.560	&	123.460 \\
\cline { 2 - 6 } & \text { Sharpe } & 0.276	&	0.533	&	-1.188	&	4.768 \\

\hline \text { Riskfree } &\text{} & 1.464	&	1.713	&	0.011	&	6.356 \\
\hline \text { Overall } &\text{Return} & 8.226	&	23.272	&	-176.220	&	126.320 \\
\hline \text { Overall } &\text{Sharpe} & 0.290	&	0.556	&	-2.446	&	4.768 \\
\hline
    \end{tabular} }
    \caption{Time series descriptive statistics}
    \label{tab:tab1}
\end{table}

Looking at the details, the table illustrates five fund examples with the annualised return (percent) and the annualised Sharpe ratio, in terms of their mean, standard deviation, minimum and maximum values. The five funds are respectively the minimum average return, 25th percentile, 50th percentile, 75th percentile and maximum average return over the period. The risk-free rate of return is the market yield on the 3-month U.S. Treasury Securities. The last two rows represent the overall average return and respectively the Sharpe ratio of all the funds used in the sample.

Within the primary ``large-cap equities" strategy that categorizes the funds, there are three major sub-strategies, namely growth, value and blend. Those adopting a ``growth" sub-strategy focuses on growth stocks listed in the United States, i.e., stocks with strong earnings growth potential, whereas ``value" sub-strategy means the investment targets value stocks, those evaluated to be undervalued. ``Blend" represents a mix of growth and value sub-strategies. The study sample of 634 funds includes all the three sub-categories mentioned above.

\subsection{Data preprocessing}
\subsubsection{Data splitting}
The time series data are split into 6 train and validation sets using the cross-validation scheme described in section \ref{cv-scheme} of Methodology. This cross-validation scheme allows for robust and reliable model selection based on their average out-of-sample performances. 

\subsubsection{Removing trend and seasonality}
We transform each time series in the train set of each cross-validation split into its stationary form by two commonly used techniques known as log transformation and differencing. Log transformation stabilizes the variance of the time series \citep{hyndman2013forecasting}. Since negative Sharpe ratios exist, an offset is added into all time series, to ensure each one of them is all positive, before applying log transformation. 

While log transformation is effective in handling time series variances, differencing can help remove changes in the level of a time series and is therefore effective in eliminating (or reducing) trend and seasonality \citep{hyndman2013forecasting}. Subsequent to log transformations, we examine each time series to see which one needs differencing and the number of differences required to transform them into stationary form. The result shows that the majority of the time series have been stationary after log transformation and 68 of them need to be applied first differencing. These 68 time series are applied first differencing in each cross-validation split and their last train observation is saved for later inverse transformation back to their original scales. 

\subsubsection{Data postprocessing}
Since each time series has been transformed prior to entering the modelling stage, the models' outputs are not in the original scale and this does not allow for a direct comparison of the models' outputs with the validation sets to obtain error metrics. Therefore, models' predictions are transformed back to the original scales following the inverse order that we apply in the preprocessing steps. In detail, predictions for those time series that are previously differenced receive inverse differencing; after this step, all time series are inverse log-transformed, and then the offset that was previously added is subtracted from each time series.

\subsubsection{Forecasting multiple outputs}
Unlike univariate statistical methods, deep learning algorithms can potentially take advantage of cross-learning, in which patterns from multiple time series are learnt to improve forecast accuracy for individual time series. In this work, we aim to compare forecasting methods for predicting 634 mutual funds, which involve 634 time series of a homogeneous group of funds. Each time series requires multi-step-ahead forecasts for the applicable forecast horizon. Therefore, we approach the problem as a multiple-output and multi-step-ahead forecasting problem, where inputs are fed into the neural networks, and after training, vectors of outputs are produced directly by the model for six, nine, 12, 18 and 24 months ahead. This strategy has been demonstrated effective by the works of \citep{taieb2012review} and \citep{wen2017multi} and also adopted by \citep{hewamalage2021recurrent}.

Supervised deep learning maps inputs to outputs, where in our time series context, inputs are fed into the model and outputs produced by the model following a sliding window approach. This sliding window scheme is also used in the work of \citep{hewamalage2021recurrent}. This approach considers each data sample as consisting of an input window of past observations of all time series that is mapped to an output window of future observations which immediately follow the input window in terms of time order relationship. In our experiments, each train set is preprocessed to form multiple sliding input and output windows with these characteristics, where each sliding window of the next consecutive time step has the same form as the previous one but is shifted by a one-time step. This scheme allows for the use of past (lagged) time series in predicting future time series in each data observation. 

As patterns are learnt from inputs to predict outputs, we believe it is necessary to set the length of input sequences to be at least equal to the length of output sequences. This would suggest that deep learning algorithms have sufficient information from the past to predict the future. The length of each input window should not be too large, since the use of too long input windows would significantly reduce the number of training samples and probably affect model performance. In our study, we set the lengths of input windows equal to the lengths of the forecast horizon plus two months, given the lengths of output windows are determined by the chosen forecast horizons. These choices balance the need for covering sufficient information in the input windows and the need to retain as many training samples as possible. 

\subsection{Training deep learning models}
The performance of deep learning models depends on a number of factors, including, among others, the right choice of the loss function, the optimization procedure within the training loop, and the outer optimization procedure that selects the best combination (or configuration) of hyper-parameters. Our training procedure emphasizes extensive hyper-parameter tuning based on careful consideration for hyper-parameter configurations. This section describes in detail these training and optimization procedures. 

\subsubsection{Loss function}
We use \textit{RMSE} as the loss function for training deep learning models. This loss function will be optimized when training mini-batches of data samples. This type of loss function has the advantage of restricting large, outlier errors, since large errors may result in really large squared errors. This loss function is also the metric that we optimize in the Bayesian optimization loops. 

\subsubsection{Hyper-parameter search setup}
Hyper-parameters refer to the parameters that models cannot learn during the training process but may have an effect on models' performance. For RNNs, the number of hidden layers, units in each hidden layer, and learning rate, etc., are important hyper-parameters. The process of identifying which hyper-parameter configuration results in optimal generalization performance is known as hyper-parameter optimization. We employ Bayesian optimization for this optimization procedure, which will be detailed in the next section. 

In our experiments, the following hyper-parameters are optimized:
\begin{itemize}
  \item Learning rate
  \item Number of hidden layers
  \item Units in a hidden layer
  \item Dropout rate for inputs
  \item Dropout rate for hidden layers
  \item Mini-batch size
  \item Whether to use batch normalization or not 
  \item If using batch normalization, placing it either before or after dropout layers
  \item Activation functions
  \item Weight decay
  \item Number of epochs 
\end{itemize}

Unlike \citep{hewamalage2021recurrent}, we consider the dropout rate as an important hyper-parameter for optimization. Dropout is an effective technique for addressing the overfitting problem in deep neural networks \citep{srivastava2014dropout}. This technique may be even more appropriate for time series forecasting, where datasets are usually relatively small compared to those in other fields such as NLP or Computer Vision. 

We also provide an option to employ batch normalization \citep{ioffe2015batch} to further regularize and improve training speed, which allows for the possible training of deeper networks. Originally, batch normalization was understood to work by reducing internal covariate shifts. Later research demonstrated this concept was a misunderstanding. The work of \citep{santurkar2018does} clarified that the effectiveness of batch normalization lay in its ability to make the optimization landscape significantly smoother. 

In the training process that uses batch normalization, this normalization is performed for each mini-batch. By using batch normalization, we can use higher learning rates, and be more relaxed about network initialization. As a regularizer, in some cases, we may not need to use dropout if batch normalization is already in use. In our work, it is, however, uncertain that batch normalization can replace the need for dropout, especially when our training data are limited. Therefore, we choose to optimize both of these hyper-parameters.  

There is also additional consideration for the order of batch normalization and dropout in actual implementation. The combination of batch normalization and dropout works well in some cases but decreases the model performance in others, and thus their order should be carefully considered \citep{li2019understanding}. To work around this issue, and to possibly take advantage of both dropout and batch normalization, in our experiments, we add another hyper-parameter that controls whether batch normalization layers are added to the model before or after dropout layers. The range of dropout rates in our experiments includes zero, which means the case of no dropout is covered. 

Another regularization method tuned is weight decay. Weight decay, also called L2 regularization, is probably the most common technique for regularizing parametric machine learning models \citep{zhang2021dive}. This technique works by adding a penalty term (or regularization term) into the model's loss function so that the learning process will minimize the prediction loss plus the penalty term. The updated loss function with weight decay is then given by: 
$$
L = Error +\underbrace{\psi \sum_{i=1}^{p} w_{i}^{2}}_{\text {L2 regularization }}
$$
Applied to our context, $Error$ denotes the root mean square error from the network outputs, $w_{i}$ represents the trainable parameters of the network with $\mathrm{p}$ the number of all such parameters. 

Another important hyper-parameter is the number of hidden layers, which controls the network depth. The use of regularization methods described above can help reduce models' overfitting, which as a result may allow for deeper networks to be trained. Therefore, we set the range of the number of hidden layers from one to five, with five hidden layers representing a quite deep network given the limited amount of training data.

We use Adam as the optimizer for the model training process. Adam is a computationally efficient gradient-based optimization method for optimizing stochastic objective functions, which is demonstrated to work well in practice and compares favorably to other stochastic optimization approaches \citep{kingma2014adam}.

In addition to the above, we choose to optimize other hyper-parameters including learning rate, mini-batch size, units which represent the dimensionality of the output space in a hidden layer, number of epochs, and activation functions. For each numeric hyper-parameter, as appropriate, we try to include a wide range of values for the search, without making the computation cost too high. For example, the learning rate ranges between 0.001 and 0.1, and the dropout rate ranges between 0.0 and 1.0. To control the computation cost within reasonable limits, we limit the range of the number of epochs to between 1 and 30.   

\subsubsection{Hyper-parameter optimization}
Prior studies have investigated different methods for executing this optimization procedure. Among these, grid search, random search and Bayesian optimization provide three alternatives for optimizing hyper-parameters. Bayesian optimization has been demonstrated to be more effective and will be chosen in our experiments. 

\subsubsection{Grid search}
The grid search method finds the optimal hyper-parameter configuration based on a predetermined grid of values. This requires careful consideration in the choice of grid values, as there might be no limit to the number of possible hyper-parameter configurations. For example, a learning rate that ranges between 0.001 and 0.1 can take countless possible values, and when combined with several other hyper-parameters, the search space would be so vast, such that covering all possible configurations in the search is impossible. To limit the choice of hyper-parameters in a grid, we can rely on expert knowledge and experience, but this cannot guarantee success in many cases, especially when there are a large number of hyper-parameters for tuning in deep learning models. 

\subsubsection{Random search}
Random search narrows down the number of possible hyper-parameters to search by selecting a random subset of all possible configurations. By narrowing down the search space, this algorithm reduces training time and has proved to be effective in many cases \citep{putatunda2019modified}. The research of \citep{bergstra2012random} and \citep{putatunda2019modified} show that random search is more effective than grid search as a hyper-parameter optimization method.

\subsubsection{Bayesian optimization}
Bayesian optimization (BO) is a modern hyper-parameter optimization technique that is effective for searching through a large search space that may involve a large number of hyper-parameters. It has demonstrated superior performance to the random search and grid search approach in a variety of settings (\citealp{wu2019hyperparameter}; \citealp{putatunda2019modified}). 

The essence of BO is the construction of a probabilistic surrogate model that models the true objective function, and the use of this surrogate model together with the acquisition function to guide the search process. Its procedure first defines an objective function to optimize, which may be the loss function, or some other function deemed more appropriate for model selection. In training models where the evaluation of the objective function is costly, the surrogate which is cheaper to evaluate is used as an approximation of the objective function \citep{bergstra2011algorithms}.  

The whole process of Bayesian optimization is sequential by nature since the determination of the next promising points to search is dependent on what is already known about the previously searched points. In addition to the surrogate model, another important component of BO is the acquisition function, which is designed to guide the search toward potential low values of the objective function \citep{archetti2019bayesian}. The acquisition function allows for the balance between exploitation and exploration, where exploitation means that searching is performed near the region of the current best points and exploration refers to the searching in the regions that have not been explored. 

The initial probabilistic surrogate model is constructed by fitting a probability model over a sample of points selected by random search or some other sampling method. In the next step, a new promising point to search is identified using the acquisition function. The objective function is then evaluated and then the probabilistic surrogate model is updated with the new information. The next step uses the acquisition function to suggest a further promising point. This process is repeated in a sequential manner until some termination condition is satisfied. 

Gaussian Process (GP) is a popular choice for the surrogate model. The GP can be understood as a collection of random variables which satisfies the condition that if any finite number of these random variables are combined, the result will be a joint Gaussian distribution \citep{archetti2019bayesian}.
.
Another surrogate model is Tree-structured Parzen Estimator (TPE). Unlike GP which models $P(y|x)$ directly, the TPE approach models $P(x|y)$ and $P(y)$ \citep{bergstra2011algorithms}, where $x$ represents hyper-parameters, and $y$ represents the associated evaluation score of the objective function. The hyper-parameter search space can be defined by a generative process, of which the TPE replaces the prior distributions of the hyper-parameter configuration with specific non-parametric densities; the substitutions become a learning algorithm that then creates various densities over the search space \citep{bergstra2011algorithms}. 

The TPE algorithm is implemented in the well-known  \textit{hyperopt} Python package in \citep{bergstra2013hyperopt}. In our experiments, we utilize this library for hyper-parameter optimization using the TPE algorithm of the Bayesian optimization framework, where the number of iterations is set to 800 for each deep learning method. All deep learning models are trained using a server with the following characteristics: 8-core CPU, 16 GB RAM and Linux Ubuntu 20.04.4. 

\subsection{Forecast accuracy measures}
In this section, we present the metrics used to compute forecast accuracy
that enables performance comparison among the models studied.

Let $X_{t}$ denote the observation at time $t$ and $F_{t}$ denote the forecast of $Y_{t}$. Then the forecast error $e_{t}=X_{t}-F_{t}$. 
Let's have $k$ forecasts and that observation of data at each forecast period.

We use the same notation as in \citep{hyndman2006another} mean $\left(x_{t}\right)$ to denote the sample mean of $\left\{x_{t}\right\}$ over the sample. Analogously, we use the median $\left(x_{t}\right)$ for the sample median.

The most commonly used scale-dependent measures are based on absolute errors or squared errors. Let $X_t$ denote the observation at time $t$ and $F_t$ denote the forecast of $Y_t$. Then the forecast error $e_t=X_t-F_t$.
Let Mean Square Error $MSE=\operatorname{mean}\left(e_l^2\right)$.

Root Mean Square Error:
\begin{equation*}
    RMSE = \sqrt{\operatorname{mean}\left(e_{l}^{2}\right)}
    \label{RMSE}
\end{equation*}

Mean Absolute Error:
\begin{equation*}
    MAE = \operatorname{mean}\left(\left|e_{t}\right|\right)
    \label{MAE}
\end{equation*}

\noindent Often, the RMSE is preferred to the MSE as it is on the same scale as the data. Historically, the RMSE and MSE have been popular, largely because of their theoretical relevance in statistical modelling. However, they are more sensitive to outliers than MAE or MDAE (Median Absolute Error).

Compared to absolute error, \textbf{ percentage errors} have the advantage of being scale-independent, and so are frequently used to compare forecast performance across different datasets.
Let defined percentage error as $p_{t}=100 e_{t} / X_{t}.$ \\
The Symmetric Median Absolute Percentage Error: \\
\begin{equation*}
    SMDAPE = \operatorname{median}\left(200\left|X_{t}-F_{t}\right| /\left(X_{t}+F_{t}\right)\right)
\end{equation*}

The problems arising from small values of $X_{t}$ may be less severe for SMDAPE. However, usually when $X_{t}$ is close to zero, $F_{l}$ is also likely to be close to zero. Thus, the measure still involves division by a number close to zero.

\textbf{Scaled errors} is defined as:
$$
q_{t}=\frac{e_{t}}{\frac{1}{n-1} \sum_{i=2}^{n}\left|X_{i}-X_{i-1}\right|}
$$
which is independent of the scale of the data. A scaled error is less than one if it arises from a better forecast than the average one-step Naive forecast computed in-sample. Conversely, it is greater than one if the forecast is worse than the average one-step Naive forecast computed in-sample (see \citealp{hyndman2006another}).

The famous scaled error is the Mean Absolute Scaled Error: 
\begin{equation*}
\displaystyle
    \operatorname{MASE}=\operatorname{mean}\left(\left|q_{t}\right|\right)
\end{equation*}
When MASE $<1$, the proposed method gives, on average, smaller errors than the one-step errors from the Naive method. If multi-step forecasts are being computed, it is possible to scale by the in-sample MAE computed from multi-step naïve forecasts.

The recent paper of \citep{kim2016new} investigated and provided practical advantages of the new accuracy measure MAAPE, the mean arctangent absolute percentage error:

\begin{equation*}
\displaystyle
    \operatorname{MAAPE}=\frac{1}{N} \sum_{t=1}^{N} \mathrm{AAPE}_{\mathrm{t}}=\frac{1}{N} \sum_{t=1}^{N} \arctan \left(\left|\frac{X_{t}-F_{t}}{X_{t}}\right|\right)
\end{equation*}

Although MAAPE is finite when the response variable (i.e. $X_{t}$ ) equals zero, it has a nice trigonometric representation. However, because MAAPE's value is expressed in radians, this makes MAAPE less intuitive.
Note that MAAPE does not have a symmetric version, since division by zero is no longer a concern.
The MAAPE is also scale-free because its values are expressed in radians.

\section{Findings and analysis}
In this section, we run and train six models for five different forecast horizons, i.e., 30 models in total. The forecast horizons include 6 months, 9 months, 12 months, 18 months, and 24 months. Among six models are two deep learning models (LSTM, GRU), three traditional statistical (ARIMA, ETS, Theta), and the Naive model as a benchmark.

The first subsection below presents the average results for five different forecast horizons using the six models and the ensemble models. The second subsection illustrates further details using the 18-month forecast horizon.

\subsection{Average of multiple forecast horizons}
\subsubsection{Comparison of deep learning vs statistical models }

\begin{table}[ht]
  \centering
\scalebox{0.8}{
$
\begin{array}{|l|c|c|c|c|c|}
\hline \text { Algorithm } & \text { MASE } & \text { RMSE } & \text { MAE } & \text { SMDAPE } & \text { MAAPE } \\
\hline \text { LSTM } & \textbf{1.510} & \textbf{0.441} & \mathbf{0 . 3 6 3} & \textbf{74.204} & \textbf{59.340} \\
\hline \text { GRU } & 1.546 & 0.453 & 0.371 & 81.481 & 59.791 \\
\hline \text { ARIMA } & 1.815 & 0.533 & 0.435 & 96.504 & 63.207 \\
\hline \text { ETS } & 1.835 & 0.535 & 0.441 & 105.028 & 65.012 \\
\hline \text { Theta } & 1.961 & 0.566 & 0.471 & 106.827 & 67.567 \\
\hline \text { Naive } & 4.176 & 1.938 & 1.437 & 120.213 & 75.317 \\
\hline
\end{array}$
}
    \caption{Average accuracy measure of five forecast horizons}
    \label{tab: Average accuracy measure of five forecast horizons}
\end{table}

Table \ref{tab: Average accuracy measure of five forecast horizons} provides the averages of the accuracy measures over five forecast horizons for each of five accuracy measures (MASE, RMSE, MAE, SMDAPE and MAAPE) and each of six models. It is evident that the LTSM outperforms all other models, then the GRU, resulting in the deep learning models taking the first place. The next places are the ARIMA model, ETS, and the Theta,  respectively. The worst model surprisingly is the Naive model. Regardless of the forecast horizon, the LSTM model is the best model on average and produces the most consistent result across different accuracy measures, as shown in Table  \ref{tab:accuracy measures across 6 algorithms
and five forecast horizons} in the appendix.

\begin{figure}[H]
    \centering
    \graphicspath{ }
\includegraphics[scale=0.5]{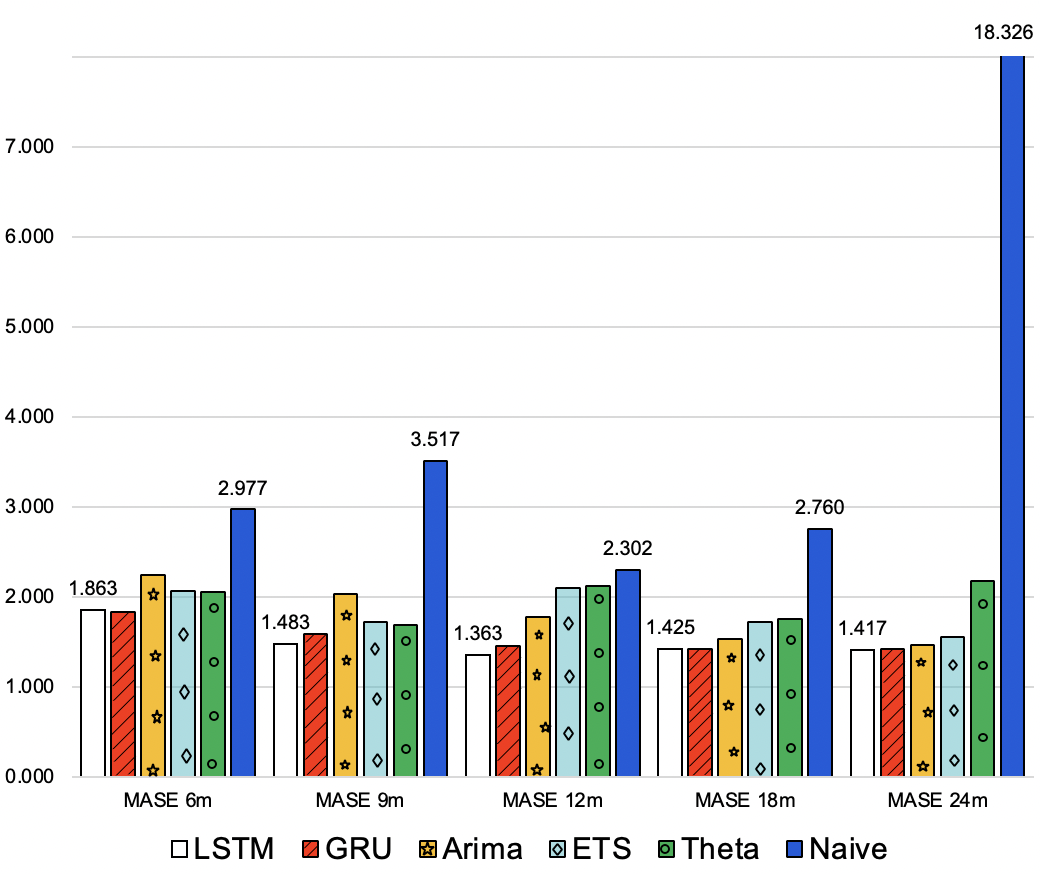}
   \caption{Comparison of MASE across all algorithms 
and forecast horizons.} 
    \label{fig:comparison}
\end{figure}

Figure \ref{fig:comparison} shows the MASE accuracy measure and RMSE and MAE are reported in Table \ref{tab:accuracy measures across 6 algorithms
and five forecast horizons} in the appendix.
The results confirm the outperformance of the LSTM model as its MASE stays lowest across six models and the different forecast horizons, except for the six-month horizon for which GRU slightly outperformed LSTM. Across the five horizons,  LSTM produces the lowest MASE (1.363) for the forecast horizon of 12 months. 

Table \ref{tab:accuracy measures across 6 algorithms
and five forecast horizons} reports RMSE, MAE, SMDAPE and MAAPE results. We conclude that LSTM is the best model across different algorithms and across different forecast horizons. Again, the 12-month horizon receives the best forecasting accuracy for the first two metrics and the nine-month for the last two.  The Naive model consistently produces the least accuracy across algorithms and forecast horizons, especially the longest one of 24 months.

\subsubsection{Ensemble models}

\begin{table}[ht]
    \centering
    \scalebox{0.75}{
 $\begin{array}{|c|c|c|c|c|c|}
\hline \text { Algorithm } & \text { MASE } & \text { RMSE } & \text { MAE } & \text { SMDAPE } & \text { MAAPE } \\
\hline & \multicolumn{5}{|c|}{\text { All algorithms }} \\    
\hline \text { simple average } & 2.208 & 0.678 &  0.530 & 94.532 & 63.984 \\
\hline \text { global weights } & 1.852 & 0.553 &  0.445 & 91.495 & 62.839 \\
\hline \text { local weights } & \textbf{1.785} & \textbf{0.529} &  \textbf{0.428} & \textbf{91.004} & \textbf{62.567} \\
\hline & \multicolumn{5}{|c|}{\text { Deep learning }} \\
\hline \text { simple average } & 1.469 & 0.431 &  0.353 & 74.315 & 58.429 \\
\hline \text { global weights } & \textbf{1.468} & \textbf{0.430} &  \textbf{0.352} & \textbf{74.261} & \textbf{58.418} \\
\hline \text { local weights } & 2.056 &   0.594 & 0.493 & 127.001 & 67.196 \\
\hline & \multicolumn{5}{|c|}{\text { Stats }} \\
\hline \text { simple average } & 2.721 & 0.841 &  0.654 & 110.821 & 68.312 \\
\hline \text { global weights } & 2.119 & 0.631 &  0.509 & \textbf{108.158} & 66.826 \\
\hline \text { local weights } & \textbf{2.074} & \textbf{0.609} &  \textbf{0.498} & 122.540 & \textbf{66.737} \\
\hline
\end{array}$}

    \caption{Average accuracy measure of ensemble method.}
    \label{tab:Average accuracy measure of ensemble method}
\end{table}

Table \ref{tab:Average accuracy measure of ensemble method} provides the results of the ensemble methods applied to combine forecasts by different groups of models: all models (All algorithms in the table), LSTM and GRU (Deep learning), and statistical models (Stats). For each ensembling group,  we use a set of combination methods including simple average, weighted average with global weights (referred to as global weights in the table), and weighted average with local weights (referred to as local weights in the table).

The best results come from the ensemble models of LSTM and GRU. The second best is All algorithms. The ensembles of All algorithms provide better accuracy for all error metrics than ensembling using only three statistical methods. Furthermore, ensembling provides lower error metrics than the individual ones for deep learning methods for simple average and global weights methods. For example, the average - mean (across five forecast horizons) of MASE for LSTM and GRU are respectively 1.510 and 1.546, whereas the highest mean of ensemble methods of simple average and global weights are only 1.469 and 1.468. The same results are also obtained for other error metrics such as RMSE, MAE, SMDAPE, and MAAPE. However, the local weights ensemble of GRU and LSTM are worse than the individual ones (true with MASE, RMSE, MAE, SMDAPE, and MAAPE), so the ensemble model with All algorithms has resulted in better accuracy measures.   

While the ensemble of All algorithms provides much better accuracy measures than the ensemble of statistical models, it is not as good as that of the original individual deep learning models. 

Initially, one would expect that the ensemble of all models should provide the best accuracy measure as, theoretically, a more diverse set of models can result in fewer correlated errors. However, as shown by our analysis, the ensembles of GRU and LSTM using global weights provide the best model with the lowest mean of all error metrics.

The relative performance of ensembling methods differs across ensembling groups and forecast horizons. For All algorithms ensembling models, across all forecast horizons (see table \ref{tab:Accuracy measures of Ensemble models}), the local weights methods provide the lowest accuracy measures. For the ensembling of the deep learning models, the best choice is the global weight for all the forecast horizons, except the 24-month horizon. For the ensembling of the Stats model, the choice is the global weight for only the 6, 9 and 12-month periods.

Further details on each forecast horizon for six original models and nine ensemble models (three ensemble combinations and three weighting methods) are provided in Tables \ref{tab:accuracy measures across 6 algorithms
and five forecast horizons} and \ref{tab:Accuracy measures of Ensemble models} in the appendix. 

The next subsection further illustrates the detailed results obtained for the forecast horizon of 18 months. Detailed analysis for other forecast horizons can be performed in a similar manner using the results provided in the appendices.

\subsection{Forecast horizon of 18 months}
\subsubsection{Comparison of deep learning vs statistical models}
For each algorithm, we first calculate the average accuracy metrics for each time series across the cross-validation splits. The results for all time series of an algorithm are then averaged to yield the final metrics' values representing the performance of that particular algorithm. For the latter step, we also report the median and standard deviation in addition to the mean.

Table \ref{tab:Accuracy measures of 6 forecasting models} presents the accuracy metrics measured for two methods of Deep learning (LSTM, GRU) and three statistical methods (ARIMA, ETS, and Theta) and the benchmark model Naive.

\begin{table}[h]
    \centering
    \scalebox{0.7}{

\begin{tabular}{|c|c|c|c|c|c|c|}
\hline Algorithm & LSTM & GRU & ARIMA & ETS & Theta & Naive \\
\hline \multicolumn{7}{|l|}{ \textbf{MASE} } \\
\hline mean & \textbf{1.425} & 1.428 & 1.539 & 1.721 & 1.762 & 2.760 \\
\hline median & \textbf{1.374} & 1.384 & 1.500 & 1.700 & 1.699 & 1.684 \\
\hline $\mathrm{sd}$ & 0.177 & 0.153 & 0.172 & 0.159 & 0.225 & 3.234 \\
\hline \multicolumn{7}{|l|}{ \textbf{RMSE} } \\
\hline mean & \textbf{0.422} & 0.422 & 0.454 & 0.503 & 0.513 & 0.830 \\
\hline median & 0.415 & \textbf{0.412} & 0.445 & 0.498 & 0.498 & 0.497 \\
\hline $\mathrm{sd}$ & 0.045 & 0.044 & 0.044 & 0.043 & 0.053 & 1.019 \\
\hline \multicolumn{7}{|l|}{ MAE } \\
\hline mean & \textbf{0.342} & 0.343 & 0.370 & 0.414 & 0.424 & 0.660 \\
\hline median & \textbf{0.334} & 0.335 & 0.363 & 0.411 & 0.411 & 0.408 \\
\hline $\mathrm{sd}$ & 0.038 & 0.033 & 0.037 & 0.034 & 0.049 & 0.765 \\
\hline \multicolumn{7}{|l|}{ \textbf{SMDAPE} } \\
\hline mean & \textbf{95.775} & 109.261 & 111.320 & 133.133 & 132.561 & 137.438 \\
\hline median & \textbf{88.499} & 108.577 & 107.999 & 130.480 & 129.424 & 131.305 \\
\hline $\mathrm{sd}$ & 20.007 & 13.229 & 19.848 & 10.339 & 10.804 & 17.906 \\
\hline \multicolumn{7}{|l|}{ \textbf{MAAPE} } \\
\hline mean & \textbf{74.979} & 76.324 & 77.763 & 81.963 & 82.685 & 85.286 \\
\hline median & \textbf{74.674} & 75.891 & 76.613 & 82.901 & 83.061 & 82.188 \\
\hline sd & 3.839 & 3.825 & 4.402 & 3.217 & 3.873 & 12.389 \\
\hline
\end{tabular}}
    \caption{Mean accuracy measures of six forecasting models of 18 months}
    \label{tab:Accuracy measures of 6 forecasting models}
\end{table}

Overall, LSTM has the best performance with the smallest mean, median and standard error for almost all accuracy metrics except for the SMDAPE metric for which the GRU model yields a lower mean and median. The results clearly show that LSTM is significantly superior to the other four forecasting methods. While ARIMA is the best-performing traditional statistical approach, it produces a much lower accuracy level than the GRU model. The metrics' median and the standard error confirm the same observation. Therefore in the following table, we present only the mean value of five different accuracy metrics. 

It can be concluded that the methods of deep learning provide significantly more accurate forecast than the traditional statistical ones, which confirms our research hypothesis.\\

\subsubsection{Ensemble models for the 18-month horizon}
\begin{table}[h]
    \centering
    \scalebox{0.7}{
    \begin{tabular}{|l|r|r|r|r|r|}
\hline \text {Algorithm}&\multicolumn{1}{l|}{\textbf{MASE}}&\multicolumn{1}{|c|}{\textbf{RMSE}}& \multicolumn{1}{l|}{\textbf{MAE}}&\multicolumn{1}{l|}{\textbf{SMDAPE}}&\multicolumn{1}{l|}{\textbf{MAAPE}}\\
\hline &\multicolumn{5}{c|}{\textbf{All algorithms}}\\
\hline
\text {simple average}& 1.535 & 0.463 & 0.369 & 111.130 & 76.021 \\
\text{global weights}& 1.513 & 0.455 & 0.363 & 110.183 & 75.772 \\
\text {local weights}& \textbf{1.469} & \textbf{0.440} & \textbf{0.353} & \textbf{109.933} & \textbf{75.565} \\

\hline &\multicolumn{5}{c|}{\textbf{Deep learning}}\\
\hline
\text {simple average} & \textbf{1.322} & \textbf{0.393} & \textbf{0.318} & 96.948 & \textbf{73.896} \\
\text {global weights} & \textbf{1.322} & \textbf{0.393} & \textbf{0.318} & \textbf{96.939} & 73.900 \\
\text {local weights}& 1.519 & 0.472 & 0.365 & 141.321 & 74.531 \\
\hline &\multicolumn{5}{c|}{\textbf{Stats}}\\
\hline
\text {simple average}& 1.794 & 0.536 & 0.431 & 131.529 & 79.995 \\
\text {global weights}& 1.756 & 0.524 & 0.422 & 130.779 & 79.70 \\
\text {local weights}& \textbf{1.639} & \textbf{0.493} & \textbf{0.394} & \textbf{142.974} & \textbf{77.915} \\

\hline 
    \end{tabular}}
    \caption{Accuracy measures of Ensemble models of 18 months}
    \label{tab:Accuracy measures of Ensemble models of 18 months}
\end{table}

The best results come from the ensemble models of GRU and LSTM, then  All algorithms ensembling and Stats ensembling groups. Furthermore, ensembling provides lower error metrics than the individual ones for deep learning methods for simple average and global weights methods. For example, the means of MASE for LSTM and GRU are respectively 1.425 and 1.428, whereas the lowest mean of simple average and global weights is only 1.322. The same results are also obtained for other accuracy measures such as RMSE, MAE, SMDAPE, and MAAPE. However, the local weights ensemble of GRU and LSTM are worse than the individual ones (true with MASE, RMSE, MAE and SMDAPE). 

\section{Conclusion}
\label{sec:sample3}

The most interesting purpose of this study is to address the challenge of forecasting the performance of multiple mutual funds simultaneously using modern deep learning approaches, with a comparison against popular traditional statistical approaches. The deep learning approaches are studied from the cross-learning perspective, which means information from various time series is used to improve predictions of individual time series, and no external features are added to the models. In addition, we use different ensemble methods to combine forecasts generated by models of traditional and modern approaches. The results show that the ranking order of model quality for the studied methods are Ensemble of deep learning models, LSTM, GRU, ARIMA, ETS, Theta, and Naive.

The results among the ensemble methods vary depending on which models are combined. The best model comes from the ensemble using weighted averages using global weights of LSTM and GRU models.

In our study, both LSTM and GRU models are trained with Bayesian optimization, a modern approach for hyper-parameter optimization that can be effective when evaluating the model for individual configurations of hyper-parameters is costly, a property particularly true for deep learning problems. In this paper, we have outlined a detailed methodology for training these deep learning models using Bayesian optimization, which we believe could be valuable for other research. 

We conclude that deep learning models and their ensembling offer promising solutions to the dilemma question of forecasting the performance of multiple mutual funds measured by Sharpe ratios. 

\section{Appendix}
\begin{table}[h]
    \centering
    \scalebox{0.85}{
\begin{tabular}{|l|l|l|l|l|r|r|}
\hline \text {Algorithm}&\multicolumn{1}{|c|}{\text{6m}}& \multicolumn{1}{l|}{\text{9m}}& \multicolumn{1}{l|}{\text{12m}}& \multicolumn{1}{l|}{\text{18m}}& \multicolumn{1}{l|}{\text{24m}}\\
\hline & \multicolumn{5}{|c|}{ \textbf{MASE} } \\
\hline LSTM & 1.863 & \textbf{1.483} & \textbf{1.363} & \textbf{1.425} & \textbf{1.417} \\
\hline GRU & \textbf{1.835} & 1.587 & 1.453 & 1.428 & 1.429 \\
\hline Arima & 2.252 & 2.035 & 1.779 & 1.539 &1.470 \\
\hline ETS & 2.073 & 1.726 & 2.104 & 1.721 & 1.552 \\
\hline Theta & 2.055 & 1.687 & 2.119 & 1.762 & 2.183 \\
\hline Naive & 2.977 & 3.517 & 2.302 & 2.760 & 18.326 \\

\hline &\multicolumn{5}{|c|}{ \textbf{RMSE} } \\
\hline \text { LSTM } & \textbf{0.516} & \textbf{0.438} & \textbf{0.414} & \textbf{0.422} & \textbf{0.416} \\
\hline \text { GRU } & 0.519 & 0.462 & 0.443 & 0.422 & 0.418 \\ 
\hline \text { ARIMA } & 0.637 & 0.619 & 0.529 & 0.454 & 0.428 \\
\hline \text { ETS } & 0.576 & 0.522 & 0.610 & 0.503 & 0.461 \\
\hline \text { Theta } & 0.574 & 0.504 & 0.613 & 0.513 & 0.628 \\
\hline \text { Naive } & 0.890 & 1.021 & 0.668 & 0.830 & 6.280 \\

\hline &\multicolumn{5}{|c|}{ \textbf{MAE} } \\
\hline \text { LSTM } & 0.447 & \textbf{0.356} & \textbf{0.328} & \textbf{0.342} & \textbf{0.340} \\
\hline \text { GRU } & \textbf{0.439} & 0.381 & 0.349 & 0.343 & 0.343 \\
\hline \text { ARIMA } & 0.539 & 0.488 & 0.427 & 0.370 & 0.353 \\
\hline \text { ETS } & 0.496 & 0.414 & 0.505 & 0.414 & 0.373 \\
\hline \text { Theta } & 0.492 & 0 .405 & 0.509 & 0.424 & 0.524 \\
\hline \text { Naive } & 0.715 & 0.847 & 0.553 & 0.660 & 4.409 \\

\hline & \multicolumn{5}{|c|}{\textbf{ SMDAPE }} \\
\hline \text { LSTM } & \textbf{50.581} & \textbf{42.983} & \textbf{60.205} & \textbf{95.775} & \textbf{121.474} \\
\hline \text { GRU } & 56.180 & 45.544 & 73.905 & 109.261 & 122.516 \\
\hline \text { ARIMA } & 71.258 & 76.036 & 99.108 & 111.320 & 124.797 \\
\hline \text { ETS } & 59.843 & 55.429 & 133.069 & 133.133 & 143.664 \\
\hline \text { Theta } & 56.510 & 53.094 & 132.822 & 132.561 & 159.146 \\
\hline \text { Naive } & 73.452 & 85.090 & 136.191 & 137.438 & 168.894 \\

\hline & \multicolumn{5}{|c|}{\textbf{ MAAPE }} \\
\hline \text { LSTM } & 43.561 & \textbf{40.625} & \textbf{56.486} & \textbf{74.979} &\textbf{81.050} \\
\hline \text { GRU } & \textbf{40.669} & 43.733 & 56.506 & 76.324 & 81.720 \\
\hline \text { ARIMA } & 46.255 & 45.339 & 63.897 & 77.763 & 82.783 \\
\hline \text { ETS } & 45.489 & 41.978 & 71.679 & 81.963 & 83.952 \\
\hline \text { Theta } & 45.335 & 42.205 & 71.980 & 82.685 & 95.627 \\
\hline \text { Naive } & 51.922 & 59.664 & 72.535 & 85.286 & 107.177 \\
\hline
\end{tabular} 
}
\caption{Accuracy measures across 6 algorithms \\and five forecast horizons}
    \label{tab:accuracy measures across 6 algorithms
and five forecast horizons}
\end{table}

\begin{table}[htbp]
    \centering
    \scalebox{0.7}{
    \begin{tabular}{|c|c|c|c|c|c|c|}
\hline & Algorithm & MASE & RMSE & MAE & SMDAPE & MAAPE \\
\hline \multirow{13}{*}{6m} & \multicolumn{6}{|c|}{ All algorithms } \\
 & simple average & 2.026 & 0.566 & 0.485 & 57.088 & 44.819 \\
 & global weights & 2.017 & 0.562 & 0.483 & 56.973 & 44.700 \\
 & local weights & \textbf{2.011} & \textbf{0.560} & \textbf{0.482} & \textbf{56.818} & \textbf{44.655} \\
 & \multicolumn{6}{|c|}{ Deep learning } \\
 & simple average & 1.737 & 0.486 & 0.416 & 49.372 & 39.781 \\
 & global weights & \textbf{1.737} & \textbf{0.486} & \textbf{0.416} & \textbf{49.378} & \textbf{39.771} \\
 & local weights & 3.026 & 0.817 & 0.725 & 118.567 & 59.910 \\
 & \multicolumn{6}{|c|}{ Stats } \\
 & simple average & 2.206 & 0.623 & 0.528 & 62.855 & 47.206 \\
 & global weights & \textbf{2.195} & \textbf{0.618} & \textbf{0.526} & \textbf{62.800} & \textbf{47.143} \\
 & local weights & 2.546 & 0.694 & 0.610 & 86.404 & 51.459 \\
 
\hline \multirow{13}{*}{9m} & \multicolumn{6}{|c|}{ All algorithms } \\
 & simple average & 1.754 & 0.524 &  0.422 & 51.380 & 43.182 \\
 & global weights & 1.715 & 0.513 &  0.412 & 50.520 & 42.522 \\
 & local weights & \textbf{1.689} & \textbf{0.506} &  \textbf{0.406} & \textbf{49.718} & \textbf{42.029} \\
 & \multicolumn{6}{|c|}{ Deep learning } \\
 & simple average & 1.526 & 0.446 &  0.366 & 44.548 & 42.095 \\
 & global weights & \textbf{1.524} & \textbf{0.446} &  \textbf{0.366} & \textbf{44.508} & \textbf{42.042} \\
 & local weights & 2.406 & 0.696 &  0.578 & 101.304 & 53.860 \\
 & \multicolumn{6}{|c|}{ Stats } \\
 & simple average & 2.040 & 0.604 &  0.490 & 65.303 & 47.650 \\
 & global weights & \textbf{1.959} & \textbf{0.582} &  \textbf{0.470} & \textbf{62.085} & \textbf{46.262} \\
 & local weights & 2.104 & 0.626 &  0.505 & 82.616 & 46.953 \\

\hline \multirow{13}{*}{12m} & \multicolumn{6}{|c|}{ All algorithms } \\
& simple average & 1.699 & 0.507 & 0.408 & 93.867 & 63.387 \\
& global weights & 1.684 & 0.503 & 0.404 & 92.321 & 63.099 \\
& local weights & \textbf{1.680} & \textbf{0.502} & \textbf{0.404} & \textbf{92.168} & \textbf{62.985} \\
& \multicolumn{5}{|c|}{ Deep learning } \\
& simple average & 1.365 & 0.419 & 0.328 & 61.133 & 55.451 \\
& global weights & \textbf{1.363} & \textbf{0.419} & \textbf{0.328} & \textbf{60.908} & \textbf{55.453} \\
& local weights & 1.948 & 0.558 & 0.468 & 125.655 & 70.640 \\
& \multicolumn{5}{|c|}{ Stats } \\
& simple average & 2.009 & 0.588 & 0.483 & 127.343 & 69.327 \\
& global weights & \textbf{2.003} & \textbf{0.586} & \textbf{0.481} & \textbf{126.859} & \textbf{69.202} \\
& local weights & 2.039 & 0.590 & 0.490 & 139.245 & 70.686 \\

\hline \multirow{13}{*}{18m} & \multicolumn{6}{|c|}{ All algorithms } \\
& simple average & 1.535 & 0.463 & 0.369 & 111.130 & 76.021 \\
& global weights & 1.513 & 0.455 & 0.363 & 110.183 & 75.772 \\
& local weights & \textbf{1.469} & \textbf{0.440} & \textbf{0.353} & \textbf{109.933} & \textbf{75.565} \\
& \multicolumn{5}{|c|}{ Deep learning } \\
& simple average & 1.322 & 0.393 & 0.318 & 96.948 & 73.896 \\
& global weights & \textbf{1.322} & \textbf{0.393} & \textbf{0.318} & \textbf{96.939} & \textbf{73.900} \\
& local weights & 1.519 & 0.472 & 0.365 & 141.321 & 74.531 \\
& \multicolumn{5}{|c|}{ Stats } \\
& simple average & 1.794 & 0.536 & 0.431 & 131.529 & 79.995 \\
& global weights & 1.756 & 0.524 & 0.422 & 130.779 & 79.70 \\
& local weights & \textbf{1.639} & \textbf{0.493} & \textbf{0.394} & \textbf{142.974} & \textbf{77.915} \\

\hline \multirow{13}{*}{24m} & \multicolumn{6}{|c|}{ All algorithms } \\
& simple average & 4.025 & 1.329 & 0.968 & 159.198 & 92.513 \\
& global weights & 2.331 & 0.729 & 0.560 & 147.479 & 88.100 \\
& local weights & \textbf{2.073} & \textbf{0.634} & \textbf{0.498} & \textbf{146.383} & \textbf{87.601} \\
& \multicolumn{6}{|c|}{ Deep learning } \\
& simple average & 1.395 & 0.409 & 0.335 & 119.573 & 80.923 \\
& global weights & 1.395 & 0.409 & 0.335 & 119.573 & 80.924 \\
& local weights & \textbf{1.382} & \textbf{0.430} & \textbf{0.332} & \textbf{148.157} & \textbf{77.038} \\
& \multicolumn{6}{|c|}{ Stats } \\
& simple average & 5.558 & 1.853 & 1.337 & 167.076 & 97.383 \\
& global weights & 2.683 & 0.843 & 0.645 & 158.265 & 91.822 \\
& local weights & \textbf{2.042} & \textbf{0.640} & \textbf{0.490} & \textbf{161.460} & \textbf{86.667} \\

\hline
\end{tabular}}
    \caption{Accuracy measures of Ensemble models}
    \label{tab:Accuracy measures of Ensemble models}
\end{table}

\section{Acknowledgement}
\noindent
We acknowledge the support from Hanoi University for providing a server for training models, Ms Hoang Anh and Mr Nguyen Ngoc Hieu for their assistance in the initial phase of the project.  

\bibliography{fundsper_forecast}

\end{document}